\RequirePackage{fix-cm}
\documentclass[smallextended]{svjour3}       
\smartqed  
\usepackage{graphicx}
\usepackage[normal]{subfigure}
\usepackage{xspace}
\usepackage{url}
\usepackage[numbers]{natbib}

\def\testmed {TESTMED\xspace}

\def\myi    {\emph{(i)}\xspace}
\def\myii   {\emph{(ii)}\xspace}
\def\myiii  {\emph{(iii)}\xspace}
\def\myiv   {\emph{(iv)}\xspace}

\begin{document}
\begin{sloppypar}
\title{The TESTMED Project Experience}
\subtitle{Process-aware Enactment of Clinical Guidelines \\through Multimodal Interfaces}

\titlerunning{The TESTMED Project Experience}        

\author{Andrea Marrella \and
        Massimo Mecella \and
        Mahmoud Sharf \and
        Tiziana Catarci
}

\authorrunning{A. Marrella, M. Mecella, M. Sharf, T. Catarci} 

\institute{A. Marrella, M. Mecella, M. Sharf, T. Catarci \at
           Dipartimento di Ingegneria Informatica Automatica e Gestionale Antonio Ruberti \\Sapienza Universit\'{a} di Roma, Rome, Italy\\
              \email{lastname@diag.uniroma1.it}           
}

\date{Received: date / Accepted: date}

\maketitle

\begin{abstract}
Healthcare is one of the largest business segments in the world and is a critical area for future growth. In order to ensure efficient access to medical and patient-related information, hospitals have invested heavily in improving clinical mobile technologies and spread their use among doctors.
Notwithstanding the benefits of mobile technologies towards a more efficient and personalized delivery of care procedures, there are also indications that their use may have a negative impact on patient-centeredness and often places many cognitive and physical demands on doctors, making them prone to make medical errors.
To tackle this issue, in this paper we present the main outcomes of the project TESTMED, which aimed at realizing a clinical system that provides operational support to doctors using mobile technologies for delivering care to patients, in a bid to minimize medical errors. The system exploits concepts from Business Process Management on how to manage a specific class of care procedures, called clinical guidelines, and how to support their execution and mobile orchestration among doctors. As a viable solution for doctors' interaction with the system, we investigated the use of vocal and touch interfaces. User evaluation results indicate a good usability of the system.
\end{abstract}

\section{Introduction}\label{sec:introduction}


Healthcare is conventional regarded as the act of taking preventative or necessary medical procedures to improve a person's well-being. Such procedures are typically offered through a healthcare system made up of hospitals and professionals (such as general practitioners, nurses, doctors, etc.) working in a multidisciplinary environment with complex decision-making responsibilities.

With the advent of advanced health information technology (HIT) and electronic health records (EHR) in the mid-2000s~\cite{health20}, hospitals started to manage and share patient information electronically rather than through paper records. This has led to a growing usage of (handwriting capable) mobile technologies and devices
able to sync up with EHR systems, thus allowing doctors to access patient records from remote locations and to support them in the delivery of care procedures.

Nowadays, it is not uncommon to encounter doctors who interact with more than one mobile device at the same time, while visiting to a patient. Notwithstanding the benefits of EHR systems and mobile technologies towards improving the delivery of care procedures~\cite{chaudhry2006systematic,cook2009impact,buntin2011benefits,lee2013impact}, there are also indications that their use may have a negative impact on patient-centeredness~\cite{Shachak2009} and often places many cognitive and physical demands on doctors, making them prone to make medical errors~\cite{NYT} and lose rapport with their patients~\cite{Booth2004,Margalit2006}. However, as Laxmisan et al. have observed in~\cite{Laxmisan2007}, multi-tasking and information transfers through EHR systems have become necessary aspects of healthcare environments, which can not be avoided entirely.

Whereas a technological solution ensuring continuity of information flow through EHR systems, supporting doctors in the execution of care procedures, and with the potential to reduce the cognitive and physical burden on doctors using mobile technologies, is desirable, to date the most of existing efforts focuses exclusively on one aspect of the foregoing requirements or a partial combination of them \cite{reichert2011}.

%

On the one hand, the Human-Computer Interaction (HCI) community has investigated how the use of \emph{multimodal interfaces} has the potential to reduce the cognitive efforts on users that manage complex activities such as the clinical ones. For example, a study by Oviatt et al.~\cite{Oviatt2004} found that ``\emph{multimodal interface users spontaneously respond to dynamic changes in their own cognitive load by shifting to multimodal communication as load increases with task difficulty and communicative complexity}''. Furthermore, recent research by Pieh et al.~\cite{pieh2014effectiveness} has shown that \emph{multimodal approaches} to healthcare deliver the most effective results, compared to a single modality on its own.

On the other hand, the Business Process Management (BPM) community has studied how to organize clinical activities in well-structured \emph{healthcare processes} and automate their execution through the use of dedicated Process Aware Information Systems (PAISs). PAISs are able to interpret such processes and to deliver to doctors and medical staff (e.g., nurses, general practitioners) relevant information, documents and clinical tasks to be enacted, by invoking (when needed) external tools and applications~\cite{Lenz2007}. However, current BPM capabilities, driven through pre-specified and automated rules sets, have successfully addressed only some parts of the lower-level administrative processes (e.g., appointment making~\cite{YAWL4healtcare}) but have made little progress into the core care procedures.

Based on the foregoing, in this paper we present the main findings of the Italian project \testmed\footnote{\testmed was a 24 months Italian project, and stands for ``\emph{meTodi e tEcniche per la geSTione dei processi nella MEdicina D'urgenza}'', in English: ``\emph{methods and techniques for process management in emergecy healthcare}''.}, whose purpose was to design and develop a clinical PAIS, referred to as \testmed system, which investigated vocal and touch interfaces as a viable solution to reduce the cognitive load of doctors interacting with (clinical) mobile devices during the patient's visit, and a process-aware approach for the automation of a specific class of care procedures, called \emph{clinical guidelines} (CGs). CGs are recommended care pathways (presented in form of ``best practices'') providing doctors with appropriate knowledge to enact medical treatments for particular patient conditions. The use of guidelines that capture both literature-based and practice-based evidence is becoming a reality in hospitals all around the world~\cite{Sonnenberg2006,Peleg2003,deClercq2004,Wang02}.

The objective of the project was not on automating the clinical decision making, but on supporting doctors in the enactment of CGs, delivering them the relevant clinical information (such as the impact of certain medications, contraindications, etc.) to reduce the risk arising from a decision.
The system exploits concepts from BPM on how to organize CGs and how to support their execution, in whole or in part.
In addition, the system supports mobile, multimodal, hands-free and eyes-free vocal interaction with the core clinical devices, and does provide alternative support for multi-touch and visual interaction. 
This allows the doctor to switch between different modes of interaction selecting the most suitable (and less distracting) one during a patient's visit.

%

The \testmed system has been developed using the traditional User Centered Design (UCD) methodology~\cite{Dix:1997} and evaluated with DEA (``Dipartimento di Emergenza ed Accettazione'', i.e., Department of Emergency and Admissions) of Policlinico Umberto I, which is the main hospital in Rome (Italy).
The evaluation was performed in the emergency room of DEA; the target was to demonstrate that the adoption of clinical mobile devices providing multimodal user interfaces coupled with a process-oriented execution of clinical tasks represents a valuable solution to support doctors in the execution of CGs.
%

%


The rest of the paper is organized as follows.
Section \ref{sec:background} provides relevant background knowledge about healthcare processes and CGs, and introduces a concrete CG that will be used to explain the approach underlying the TESTMED system.
Section \ref{sec:approach} describes the general approach used for dealing with the enactment of CGs, while Section \ref{sec:system} presents the architecture of the system, introducing technical details of its software components.
Then, Section \ref{sec:validation} presents the outcomes of the user evaluation of the system and some performance tests. Finally, Section \ref{sec:related_work} discusses relevant works and Section \ref{sec:conclusion} concludes the paper by 
tracing future work. 
\section{Background}
\label{sec:background}

\subsection{Healthcare Processes}
\label{sec:care_processes}

Generally, in hospitals, the work of the medical staff is burdened by many organizational and clinical tasks. Care procedures must be planned and prepared, and results be obtained and evaluated. In addition, various organizational units are involved in the treatment process of a patient. For example, for a patient treated in a department of internal medicine, tests at the laboratory and the radiology department may become necessary, doctors from other units may need to come and see the patient, and reports have to be written, sent, and evaluated. Thus, all clinical tasks must be performed in certain orders, and cooperation between organizational units as well as the medical staff is required to properly achieve such tasks \cite{reichert2011}.

Based on the foregoing, several \emph{healthcare processes} of different complexity and duration can be identified. One can find short organizational procedures like order entry and result reporting for radiology, but also complex and long-running (even cyclic) treatment processes like chemotherapy.
According to~\cite{mans2015process}, healthcare processes can be organized into two main abstract classes: \emph{elective care} and \emph{non-elective care}.
\begin{itemize}
\item \emph{Elective care} relates to care for which it is medically sound to postpone treatment for days or weeks~\cite{lillrank2004standard}. According to~\cite{gupta2008appointment}, elective care can be classified into three subclasses: \myi \emph{standard processes}, for which a standardized treatment path exists which defines the different activities in the process and their timing; \myii \emph{routine processes}, where the overall outcome of the process is usually known, but different process paths may be followed during treatment; and \myiii \emph{non-routine processes}, where the next step of the process depends on the patient's reaction to an individual treatment~\cite{vissers2005health}.
\item \emph{Non-elective care} refers to \emph{emergency care}, which has to be performed immediately, and \emph{urgent care}, which can be postponed for a short time.
\end{itemize}

\begin{figure}
\centering
\includegraphics[width=0.95\columnwidth]{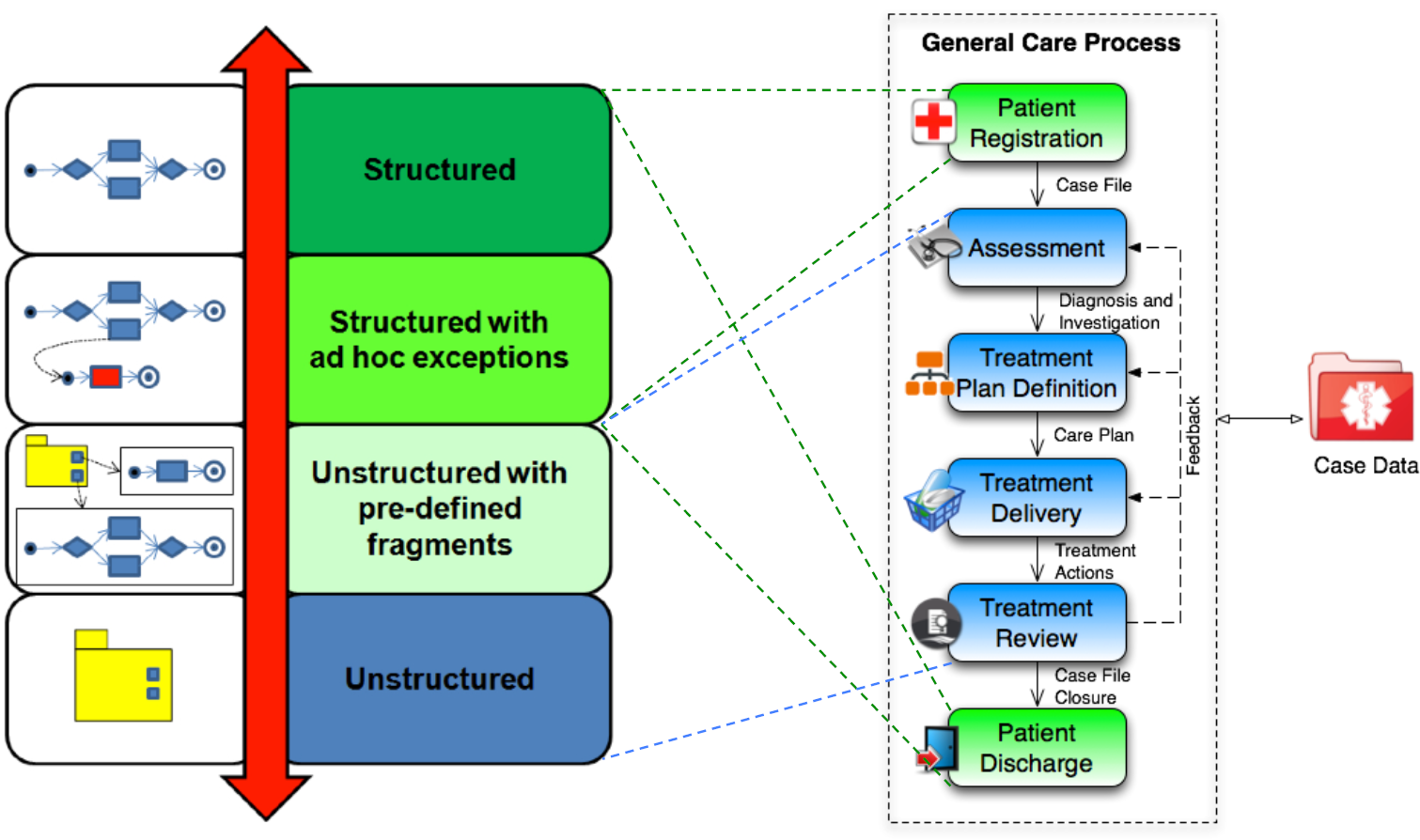}
\caption{Classification of a healthcare process~\cite{IJBPIM2013}}
\label{fig:spectrum}
\end{figure}

The complexity of a healthcare process can be additionally understood by classifying its main macro-steps along a spectrum on the basis of the degree of structuring and predictability they exhibit~\cite{IJBPIM2013}, as shown in Figure~\ref{fig:spectrum}. At the highest level of abstraction, a general healthcare process encompasses six macro-steps~\cite{acmbookhealth} including: \emph{(i)} \emph{patient registration}, resulting in the creation of the current medical case file, \emph{(ii)} \emph{patient assessment}, resulting in an initial diagnosis and in specific required investigations, \emph{(iii)} \emph{treatment plan definition}, resulting in the development of an individual care plan, \emph{(iv)} \emph{treatment delivery}, resulting in treatment actions performed according to the care plan, \emph{(v)} \emph{treatment review}, resulting in a continuous evaluation of treatment impact and efficacy that provides feedback for the previous steps, and \emph{(vi)} \emph{patient discharge}, resulting in the closing of the case records.

Administrative and organisational steps, including patient registration/discharge and other activities in the treatment delivery stage (e.g., patient transfer, bookings, management of prescriptions and lab tests) are typically structured, relatively stable and repetitive, and represent a good setting for the application of traditional approaches for process automation and improvement \cite{reichert2011}. Exceptional behaviours are limited and can be often anticipated and managed according to predefined handling procedures.

Conversely, the diagnostic and therapeutic steps driven by clinical decision-making and case data are clearly knowledge-intensive activities that lead to loosely structured or unstructured processes~\cite{JODS2014}. Clinical decision-making is highly knowledge-driven, as it depends on medical knowledge and evidence, on case- and patient-specific data, and on doctors' expertise and experience. Patient case management is mainly the result of knowledge work, where doctors act in response to relevant events and changes in the clinical context on a per-case basis, according to so-called diagnostic-therapeutic cycles based on the interleaving between observation, reasoning and action \cite{Lenz2007}.

The overall healthcare process, even in the oversimplified view provided in Figure \ref{fig:spectrum}, reflects the combination of predictable and unpredictable elements. In practice the actual flow of work in a healthcare environment may include many concurrent activities and procedures, especially in the (common) case of patients treated for multiple conditions, leading to multiple interacting care pathways.
In addition, many complicating circumstances -- often not easily predictable in advance -- may arise during the enactment of healthcare processes, making extremely complex their automation through traditional Process-aware Information Systems (PAISs), which tend to restrict too much the range of actions of medical staff~\cite{IJBPIM2013}.

Whereas it is evident that exists a gap between technology-driven approaches developed by the BPM community and methodological-based approaches suggested in the medical informatics field that is unlikely to be solved in near future \cite{reichert2011}, in this paper we discuss how a process-aware approach can be efficiently used to support the management of a specific class of care procedures, called \emph{clinical guidelines}.

\subsection{Clinical Guidelines}
\label{sec:clinical_guidelines}

In the last decades, the medical community has been actively investigating, developing and promoting evidence-based \emph{clinical guidelines} (CGs) and care pathways, as a mean for standardising clinical practice and reduce errors and costs, while improving quality of care and patient outcome ~\cite{Peleg2003,deClercq2004,peleg2009design}.
CGs are systematically developed statements to assist practitioners and patient decisions about appropriate health care for specific clinical circumstances~\cite{clinicalguidelines}. CGs are based on the best available medical research evidence, and provide advice on clinical best practices in the form of evidence-based recommendations to support and facilitate appropriate decision making in patient care.

A CG may thus provide a high-level plan of suggested/expected care and serves as a reference framework for evaluating clinical practice, but usually does not define mandatory requirements. CGs capture domain-specific knowledge but they are not defined to be directly applied to a specific patient in a particular healthcare organisational context, and need to be adapted to obtain concrete medical pathways.

\begin{figure}[t!]
\centering
\includegraphics[width=0.95\columnwidth]{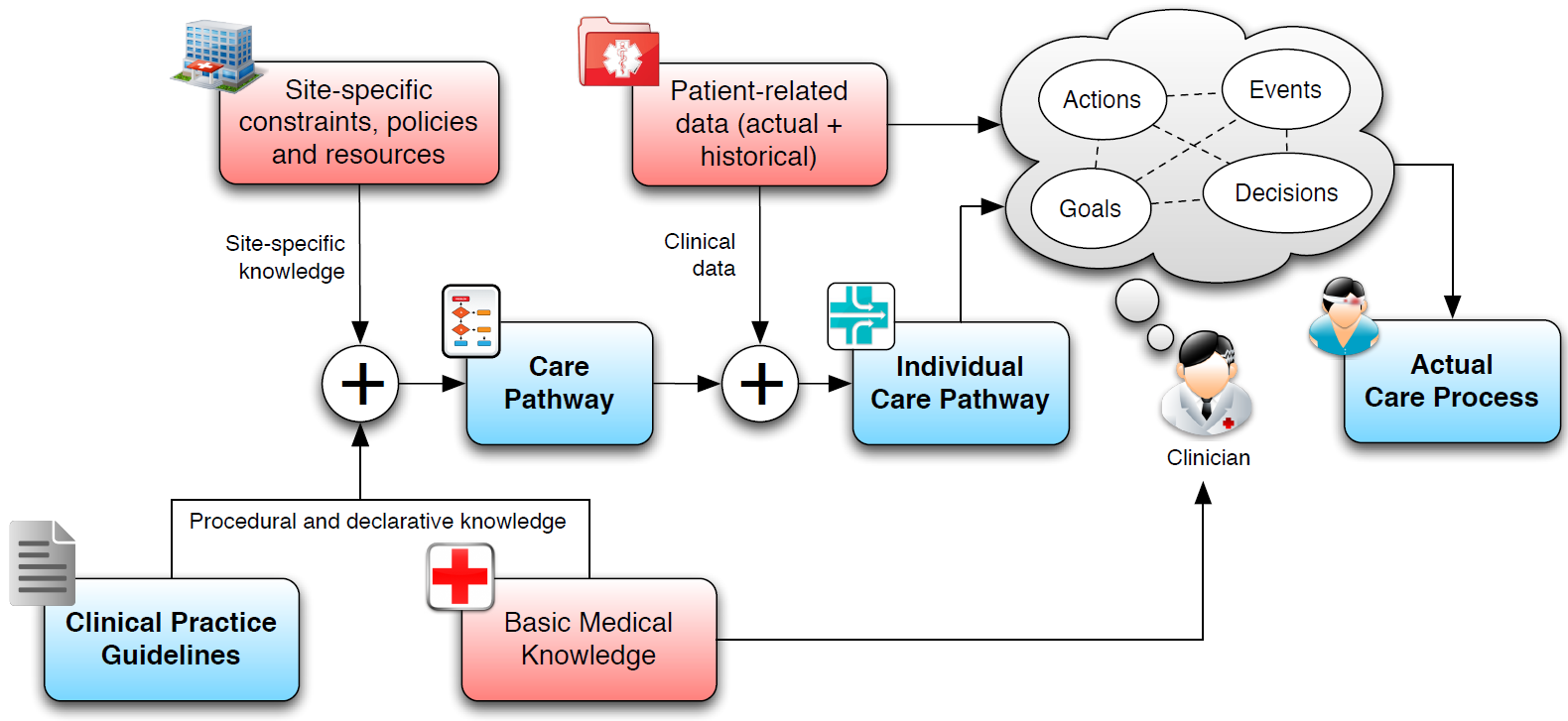}
\caption{From CGs to the actual healthcare process.}
\label{fig:CGs-2-actual}
\end{figure}

As shown in Figure~\ref{fig:CGs-2-actual}, in order to be effectively exploited in practice, the evidence-based knowledge provided by CGs has to be complemented by additional ``knowledge layers'' that include doctors' basic medical knowledge (BMK), site-specific knowledge and patient-related information, such as current conditions and medical history. Care pathways thus represent site-specific implementations of CGs. Care pathways are structured multidisciplinary care plans that describe the tasks to be carried out together with the timing, sequencing and role constraints for these tasks~\cite{carepathways}. They provide detailed guidance for each stage in the management of a patient, on the basis of intermediate and long term expected outcomes and goals. Care pathways can be considered as templates and, as detailed in \cite{Lenz2007}, the combination of care pathways and patient-related information leads to the definition of an \emph{individual patient treatment plan} that is finally carried out resulting in the actual patient treatment process.


\emph{Although the knowledge-intensive nature of clinical decision making leads to loosely structured or unstructured working procedures, the adoption of CGs and pathways introduces a process-oriented perspective in the management of patient care}. Process and decision support for patient management has been investigated in the medical informatics community through the development of models, languages and systems for the specification and execution of CGs and care pathways. Over the years, many research groups have focused on so-called ``computer-interpretable clinical guidelines'' (CIGs)
and different languages have been proposed for encoding, managing and executing CGs (see~\cite{IJBPIM2013} for a recent survey),
%
such as GLIF, Asbru, EON, PROforma, GUIDE, Prodigy and GLARE. Such languages can be broadly classified as rule-based (e.g., Arden Syntax), logic-based (e.g., PROforma), network-based (e.g., EON) and workflow-based (e.g., GUIDE). In addition, most of them are supported by systems that allow the definition and enactment of CGs~\cite{cigsystemsreview}.

Despite the availability of different formalisms, none of them has emerged over the others. Most of the existing languages are based on a process and activity-centric approach, and provide support for representing the procedural knowledge contained in CGs mainly focusing on the control-flow dimension. CGs are modeled as so-called task networks, where modeling primitives for representing actions/tasks and decisions are linked via scheduling and temporal constraints, often in a rigid flowchart-like structure. However, the efforts required to tailor and adapt CG models to specific medical settings and changing conditions are among the main barriers to their uptake. This has made the automated enactment of CGs using PAISs and process-oriented approaches as a relevant and timely challenge for the medical community~\cite{isern2008computer,reichert2011,IJBPIM2013}.

In this paper, we tackle this challenge by presenting the main findings of the TESTMED project, whose aim was to realize a clinical PAIS able to interpret CGs and orchestrate their execution among doctors and medical staff through mobile technologies and multimodal user interfaces.

\subsection{Case Study: Chest Pain}
\label{subsec:Chest_Pain}

In order to have a better understanding of the TESTMED project, we discuss the standard CG enacted for patients suffering from chest pain\footnote{Chest pain is defined as a pain that ranges from the base of the nose and navel and between the neck and the twelfth vertebra and that has no clearly identifiable traumatic cause.}, which is one of the most common reasons for the admission in the emergency room (5\% of all visits) with high mortality in case of failure diagnosis and improper dismissal (2--4\%)\cite{chestpain}.

\begin{figure}
\centering
\includegraphics[width=0.80\columnwidth]{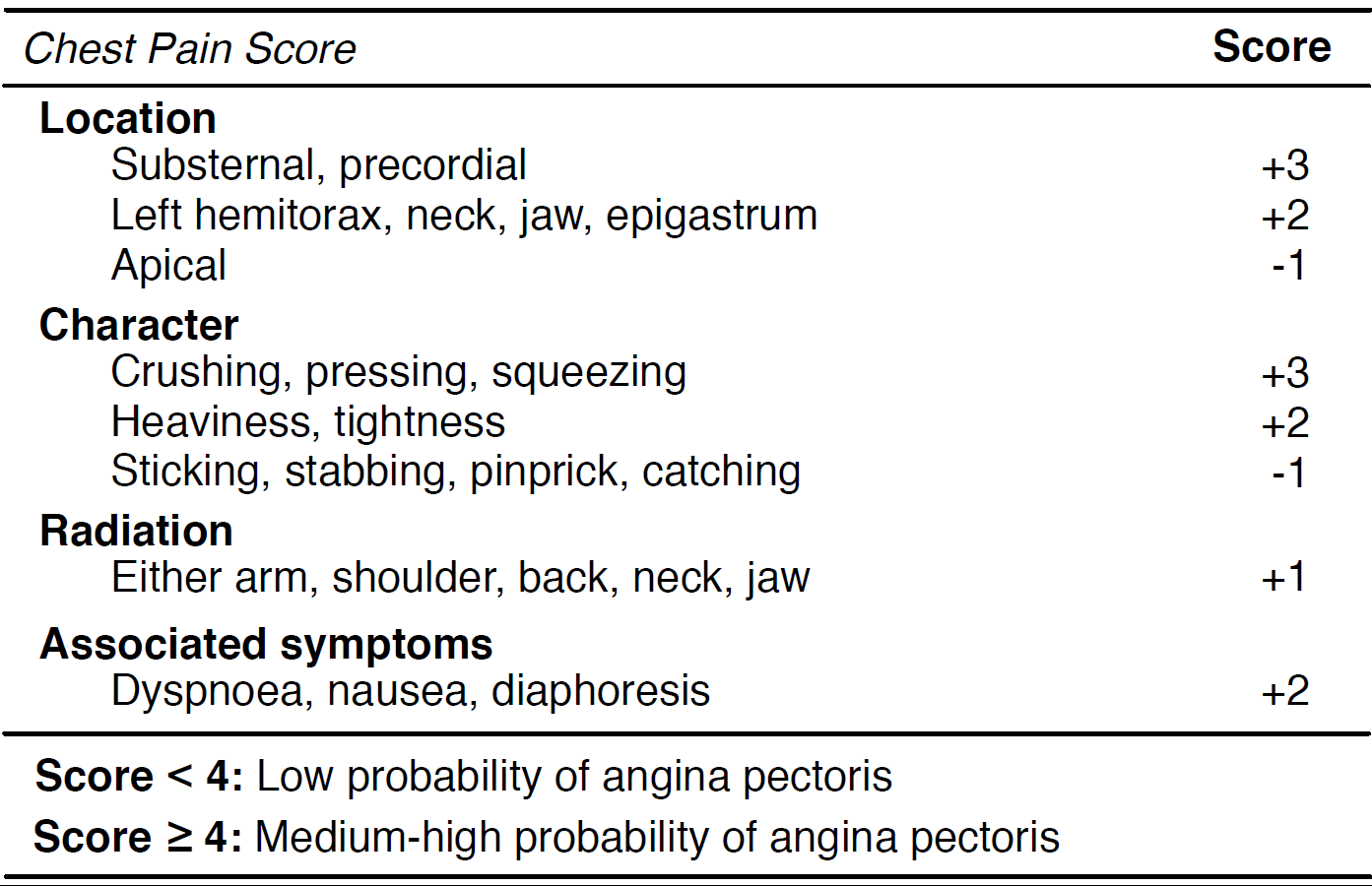}
  \caption{Chest pain score~\cite{cps}}
  \label{fig:chestpain}
\end{figure}

Typically, a patient suffering from chest pain is checked by a resident on duty in the emergency room and, on the basis of general impressions, patient history, risk factors and the so called \emph{chest pain score}, a decision is taken whether or not to admit the patient for clinical observation. The chest pain score allows to assess the clinical characteristics of acute chest pain, by calculating a semi-quantitative score. The score is used to improve the diagnostic and prognostic accuracy, in order to safely classify patients into low and high-risk subsets for cardiac events.

Figure~\ref{fig:chestpain} depicts the original scores adopted in the DEA. The score is derived by evaluating a set of four clinical characteristics:  \myii the \emph{localization} of the pain; \myii the \emph{character} of the pain; \myiii the \emph{radiation} of the pain and the \myiv \emph{associated symptoms}. A partial score is associated to every characteristic, and the sum of these values produces a final score that predicts the angina probability. A chest pain score lower than 4 identifies a low-risk probability of coronary disease, whereas a score greater or equal than 4 can be classified as an intermediate-high probability of coronary risk. Different values of the rate correspond to different clinical treatments to be followed by the patient.  
\section{Enactment of Clinical Guidelines with TESTMED}
\label{sec:approach}

The main challenge tackled by the TESTMED project was to reduce the gap between the fully automated solutions provided by the BPM community and the clear difficulties of applying a traditional process management approach in the healthcare context. To realize this vision, the major outcome of the project was the development of a clinical PAIS, referred to as TESTMED system, enabling the interpretation and execution of CGs and their presentation to doctors and medical staff through multimodal user interfaces.


The \testmed system is thought to be used when a patient suffering from a medical condition (amenable to a CG) asks for a visit. The doctor on duty in the emergency room is equipped with a tablet PC that runs the TESTMED system. The tablet PC supports multimodal (tactile and vocal) interaction, and enables the doctor to select, instantiate, and carry out specific CGs.

\begin{figure}[t!]
\centering
  \subfigure[Screenshot of an intermediary page of the questionnaire]{\includegraphics[width=0.85\columnwidth]{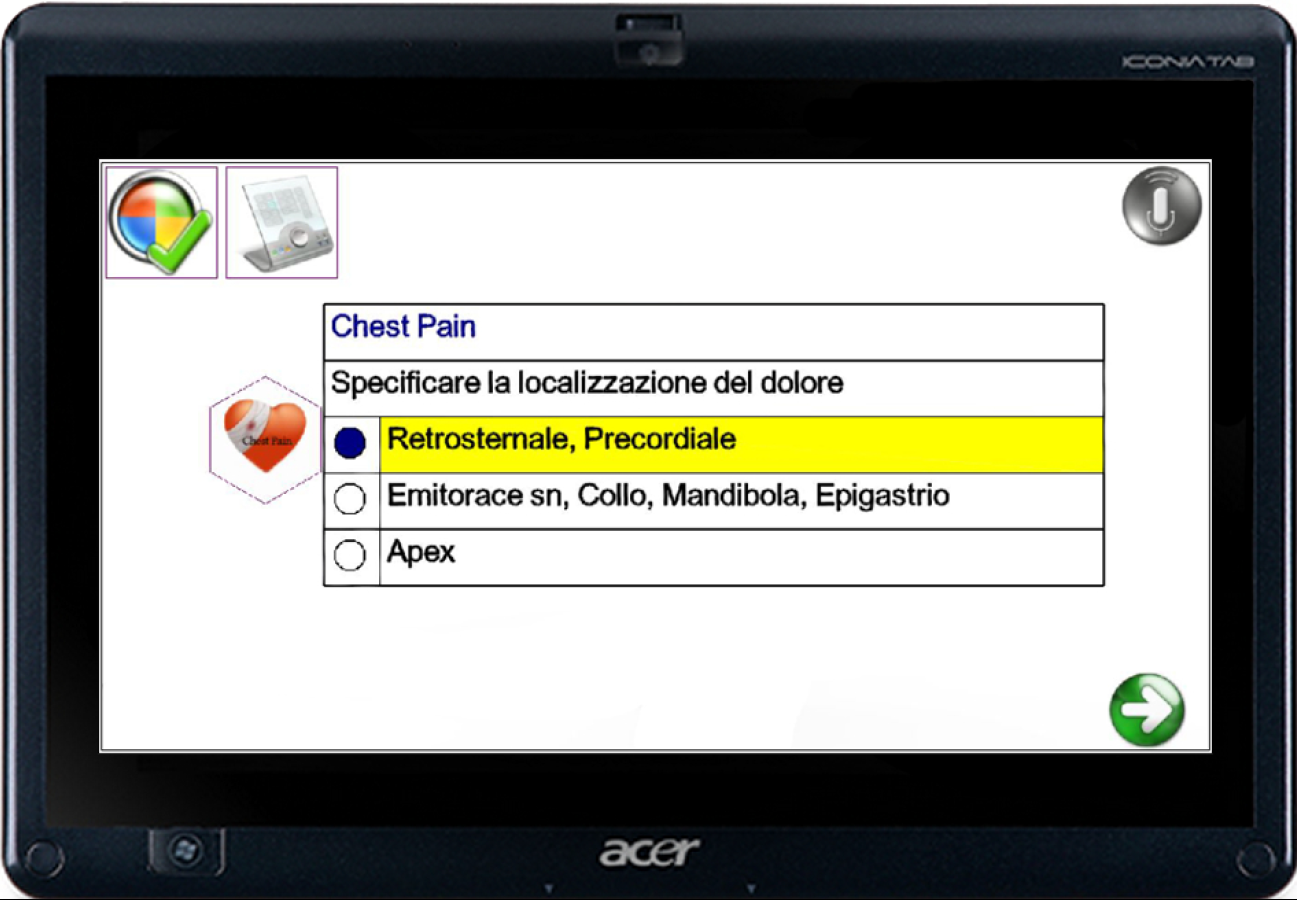}}
   \subfigure[Notification of a lab analysis result]{\includegraphics[width=0.85\columnwidth]{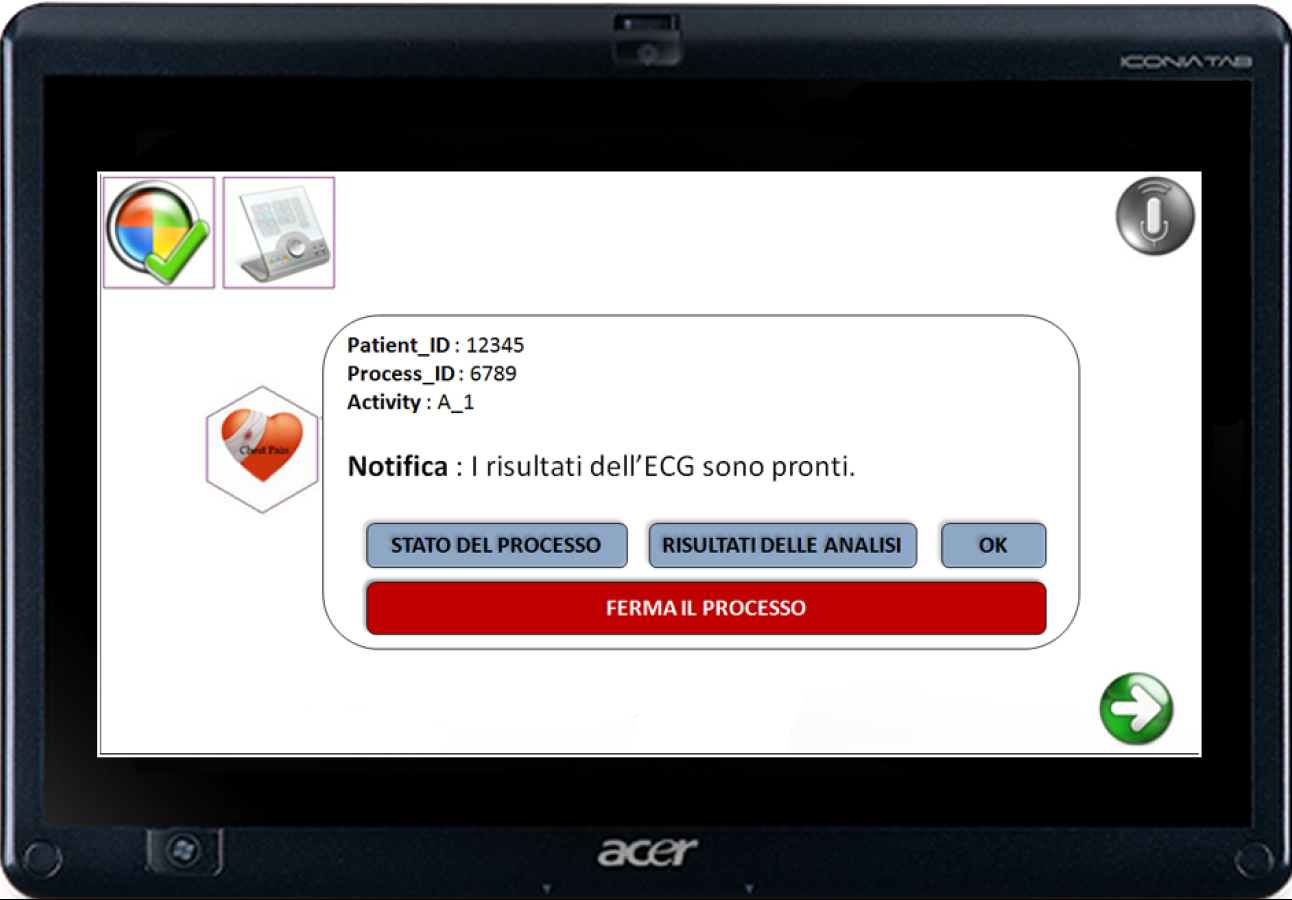}}
  \caption{The vocal/touch user interface}\label{fig:gui_two}
\end{figure}

For example, in the case of chest pain, the doctor starts filling a survey for determining the severity of the patient's medical condition, which is expressed through a chest pain score (cf. also Section \ref{subsec:Chest_Pain}). The use of the TESTMED system allows presenting the survey to the doctor both vocally, through integrated speech synthesis and recognition, and in a textual form through the GUI of the tablet PC (see Figure~\ref{fig:gui_two}(a)).
%

To exploit vocal interaction, the doctor wears a headset with a microphone linked to the tablet; s/he can listen to the questions related to the survey and reply vocally by choosing one of the speech-synthesized possible answers. Each answer is coupled with a specific characteristic and provides an associated rate.
Vocal interaction ensures that the doctor's eyes and hands are free to attend to the patient. Moreover, since the device is mobile, the doctor can move about attending to the patient and can also have mobile access to information.

\begin{figure}[t!]
\centering
\includegraphics[width=0.9\textwidth]{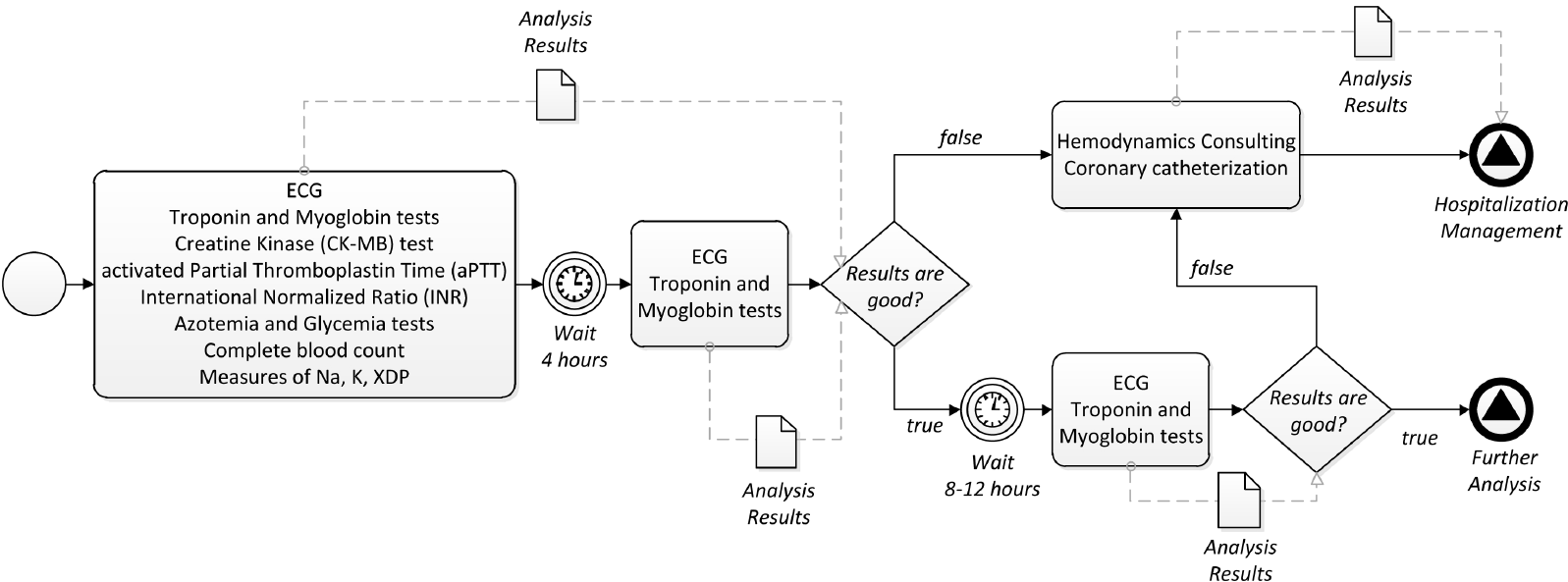}
  \caption{The care pathway for a score greater than 4 represented as a BPMN process.}
  \label{fig:process}
\end{figure}

After the survey completion, the system proposes - in the form of a care pathway - a therapy consisting of a list of medical treatments and analysis prescribed to the patient. For example, if the chest pain score of a patient is greater than 4, the suggested care pathway is similar to the one shown in Figure~\ref{fig:process}. Here the care pathway is modeled through the \emph{Business Process Modeling Notation} (BPMN\footnote{BPMN is a standard (ISO/IEC 19510:2013, cf. https://www.iso.org/standard/62652.html) to model business processes.}). The reader should notice that BPMN is not the notation employed to concretely represent and encode a CG in the TESTMED system (to this aim, we used PROforma language~\cite{PROFORMA}, as explained in Section \ref{sec:system}), but it is used here to show (in a comprehensive way) how care pathways usually look like.

The activities depicted in Figure~\ref{fig:process} concern, first of all, the enactment of some general analysis for the patient (e.g., ECG, complete blood count, etc.). After 4 hours from the first set of analysis, it is required to repeat some medical tests, like ECG and Troponin and Myoglobin tests. When the results of the analysis are ready, it is required to decide whether to hospitalize the patient or not. If the analysis outcomes present some values considered dangerous by the doctor, the care pathway suggests to make further tests (in this case, an hemodynamics consulting and a coronary catheterization) and, based on the results obtained, to activate a further procedure concerning the hospitalization of the patient. If the analysis outcomes are considered good by the doctor, the same medical tests (ECG and Troponin and Myoglobin tests) are repeated after further 8-12 hours. If the outcomes are again good, the process suggests to proceed with new analysis prescribed by the doctor; on the contrary, bad results mean to make an hemodynamics consulting, a coronary catheterization and to activate the procedure concerning the hospitalization of the patient.

\begin{figure}[t!]
\centering
\includegraphics[width=0.9\columnwidth]{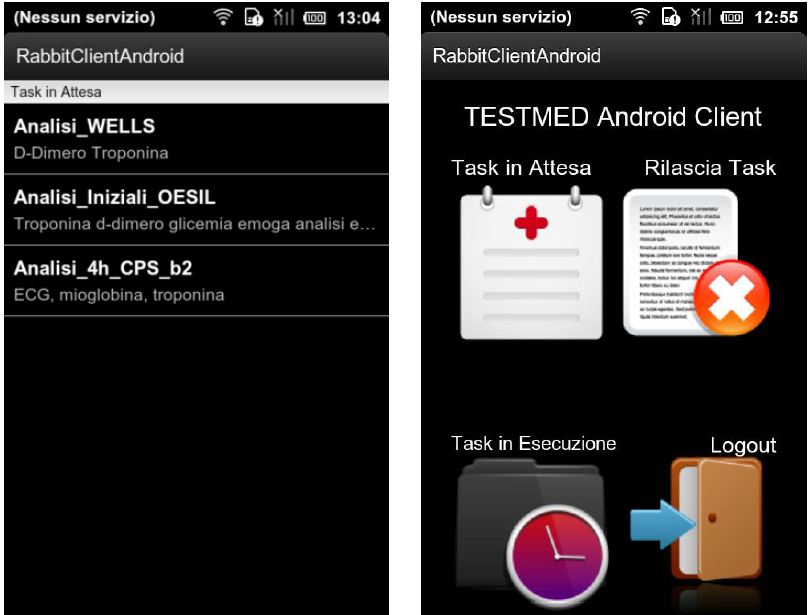}
\caption{The graphical user interface for the medical staff.}
\label{fig:device}
\end{figure}

The enactment of the various clinical tasks takes place in different moments of the therapy. Furthermore, a collaboration between doctors and medical staff is crucial to enact the proper medical treatments for each patient. The medical staff (e.g., nurses, general practitioners) are equipped with Android-based mobile devices and are notified of the progress of care pathways and of the various clinical tasks that have to be carried out for supporting doctors (e.g., to make a specific analysis, to administer a medicine, etc.). Figure \ref{fig:device} reports a couple of screenshots of the GUI provided to medical staff, which only allows for tactile interaction.

The \testmed system is able to properly orchestrate the clinical activities, assigning them to (available) doctors or members of the medical staff, and to trace the current status of the care pathway, by recording analysis outcomes and doctors' decisions. Reminders and warnings notify if new information is available for some patient (for example, if an analysis is ready to be evaluated - see Figure~\ref{fig:gui_two}(b)). In such a case, the doctor can decide to visualize further details about the analysis results and the execution status of the care path or simply to accept the notification. If there is any doubt about the goodness of the care pathway for a specific patient, the doctor can abort the process in any moment.

\section{The System Architecture}
\label{sec:system}

\begin{figure}
\centering
\includegraphics[width=0.99\columnwidth]{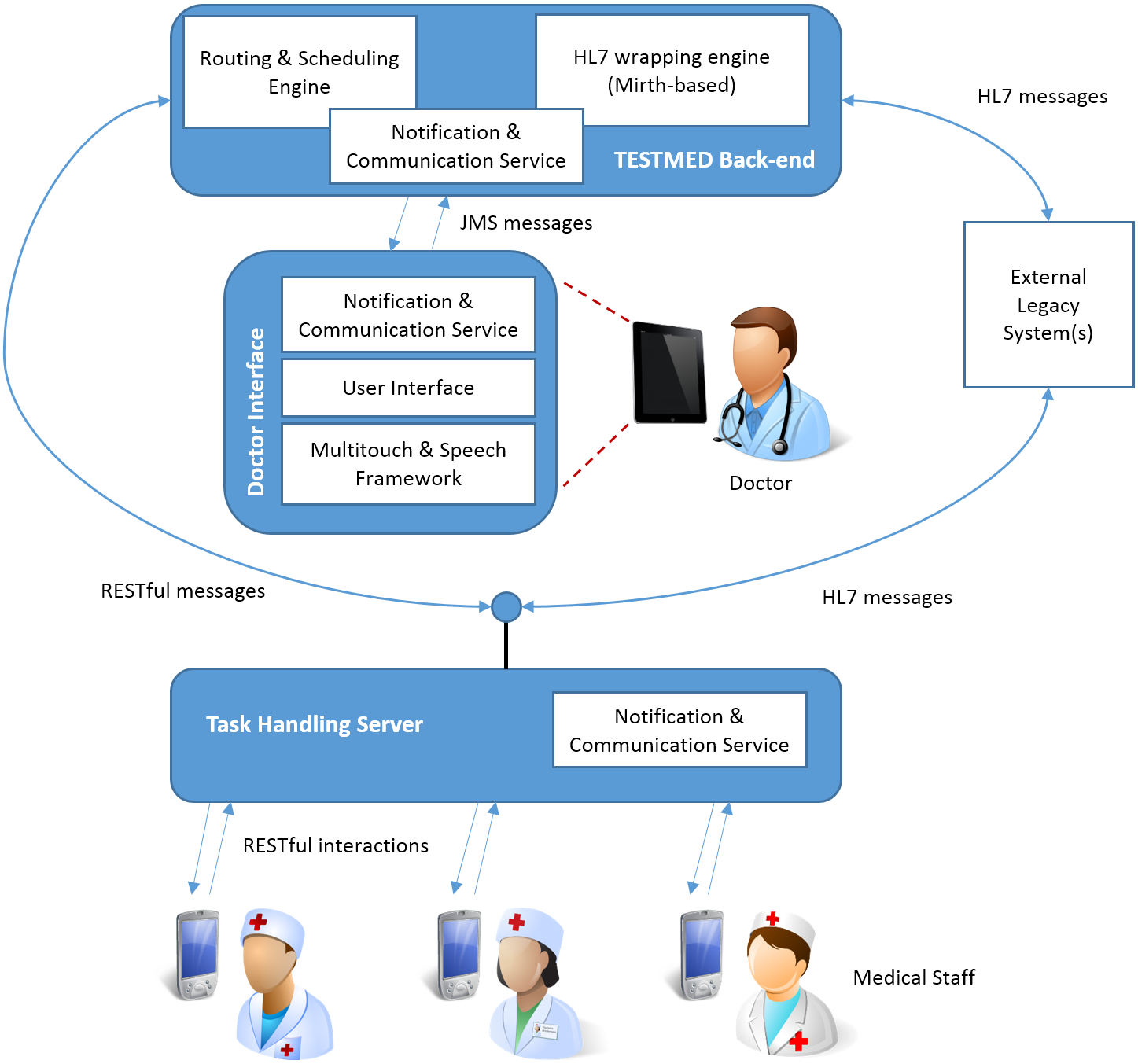}
\caption{TESTMED system architecture.}
\label{fig:arch}
\end{figure}

The \testmed system is based on three main architectural components: a \emph{graphical user interface}, a \emph{back-end engine} and a \emph{task handling server}. The overall view of the system architecture is shown in Figure \ref{fig:arch}.

The system supports mobile, hands-free and eyes-free interaction with clinical devices. %
On the one hand, doctors interact with a GUI that is specifically designed for being executed on large mobile devices (e.g., tablets), and allows for tactile or vocal interaction. In particular, vocal interaction enables doctors to work in situations where their hands and eyes are occupied with the patient's visit.
On the other hand, the GUI provided to members of the medical staff is thought to be visualized on small mobile devices (e.g., smartphones) and provides only a tactile interaction.

The \emph{back-end engine} provides the run-time environment for interpreting, activating, executing and monitoring CGs and relevant data between doctors and the medical staff. In TESTMED, a CG is specified through a combination of different languages. First, the PROforma language~\cite{PROFORMA} is used to model a guideline as a set of tasks and data items, and the control flow between them. Then, starting from the PROforma model, additional XML-based configuration settings need to be specified, to allow multimodal interaction and enable integration between system components. In such a way, a CG is finally defined as a \emph{guideline bean} and deployed into the system for being later executed.

The execution of CGs is supported by properly routing data, events and clinical tasks, according to a process-aware and content-based approach where activity scheduling and message dispatching are event- and data-driven. The interaction between all involved components and services is guaranteed by the back-end engine that manages the routing of clinical activities, relevant data, and generated events among the different actors, services, and applications. Each software component that interacts with the engine is considered as an external service to be invoked when needed.
In particular services are wrappers over pre-existing legacy systems, such as the Electronic Medical Record
(EMR) systems employed in hospitals.

The routing engine relies on a scheduler component for the timely execution of activities with temporal constraints (e.g., examinations and diagnostic laboratory tests that have to be scheduled and performed within specific time intervals), and interacts with the enterprise EMR system to \myi access and retrieve clinical and administrative patient data, \myii schedule and manage examinations, lab tests, drug prescriptions, etc. according to the clinical process, and \myiii receive events and notifications about test results and examination findings to be routed and delivered to the doctors. The interoperability with the EMR system is achieved exploiting the Health Level 7 (HL7) standard protocol\footnote{HL7 is a set of international standards for transfer of clinical and administrative data between hospital information systems. \url{http://www.hl7.org/}} and the interpretation, processing and generation of HL7 messages is managed by a specific HL7 component.

All the activities performed during a CG are supposed to be logged and recorded, in order to keep track of the events, tasks and data that contributed to the clinical and decision making process. Recorded information can be potentially exploited for: \myi reporting and analysis purposes; \myii documenting the specific patient care management process, which may have a legal relevance in case of issues; \myiii providing a knowledge base, consisting of all the patients' cases, on which to execute subsequent analysis in order to infer more evidence and possibly improving the CG itself; \myiv providing valuable support for forensic analysis~\cite{chestpainforensics}.




From a technical perspective, the multimodal interaction support is achieved by integrating the Multi-touch for Java framework (MT4j\footnote{\url{http://www.mt4j.org/}}) with the Microsoft Automatic Speech Recognition (ASR) and Text-To-Speech (TTS) engines. The back-end is realized using the Tallis engine\footnote{\url{http://archive.cossac.org/tallis/Tallis_Engine.htm}}, which has been complemented with other components for managing the integration with existing legacy systems deployed in the hospital Policlinico Umberto I. All these components are J2EE-based and hosted on a TomEE\footnote{\url{http://tomee.apache.org/apache-tomee.html}} application server. In particular, a JMS-based notification engine, namely RabbitMQ\footnote{\url{http://www.rabbitmq.com/}}, is used to manage the interaction between the doctor GUI and the back-end. The integration with the legacy systems is performed via HL7 messages over Mirth\footnote{\url{http://www.mirthcorp.com/products/mirth-connect}}.

Both the legacy systems and the back-end interact with a \emph{task handling server} via HL7 and RESTful messages. The server is in charge of notifying the medical staff about the clinical activities to be enacted for progressing the CG execution. Each member of the medical staff is equipped with a mobile device providing a specific Android client application, which interacts with the task handling server through RESTful services. 
\section{User Evaluation}
\label{sec:validation}

The \testmed system has been thought to be used in hospital wards for supporting doctors in the execution of CGs. In this context, doctors and medical staff need to collaborate in order to enact the proper medical treatments for each patient. Thus, mobile interaction is crucial, as it allows to:
\begin{itemize}
\item support the mobility of doctors for visiting the patient;
\item facilitate the continuity of information flow by enabling instant and mobile access to information;
\item expedite the doctor's decision making.
\end{itemize}
The latter point is also confirmed by a survey carried out by the Price Waterhouse Coppers' Health Research Institute (HRI) \cite{HRI2010}, which reported that 56\% of doctors indicated that mobile interaction expedited their decision making.

Whereas the use of mobile devices and applications is valuable for the improvement of collaboration and coordination amongst doctors and medical staff, there are also risks in their usage; for example, most of the clinical tasks could be highly critical and time demanding, and the challenge concerns in developing a GUI that presents relevant information in a condensed yet understandable way and captures the users' attention onto the system only when it is strictly required.

The development of specific interaction principles has required the use of \emph{user-centered design} (UCD) techniques~\cite{Dix:1997} during the life cycle of the \testmed project. Such methodologies rely on a continuous involvement of users in each phase of the project, by guaranteeing that the final system may meet user expectations. To this end, we initially produced two mockups of the system (during months 4 and 9 of the project, respectively). Each mockup was evaluated through a wide range of usability tests (controlled experiments, thinking aloud techniques, etc.) with real doctors, and the outcomes have been used for an incremental improvement of the design. For example, despite users appreciated the touch interface provided in the first mockup, they asked for an interaction with the system still less invasive. To match such a request, the vocal interface (which can be used in addition to the touch interface) has been concretely introduced in the second mockup.

Based on the above user feedbacks, we have progressively and iteratively produced two working prototypes of the system in months 12 and 18 of the project, respectively. We assessed them employing well-established usability evaluation methods involving the target users (i.e., real doctors and members of the medical staff). Results and findings of the user evaluation are discussed in the next section.

\subsection{Evaluation Setting and Results}
\label{sec:results}

The two developed working software prototypes have been tested by deploying on the system the CG enacted for patients suffering from chest pain and discussed in Section \ref{subsec:Chest_Pain}.

\begin{figure}[t]
\centering
\subfigure{\includegraphics[width=0.49\columnwidth]{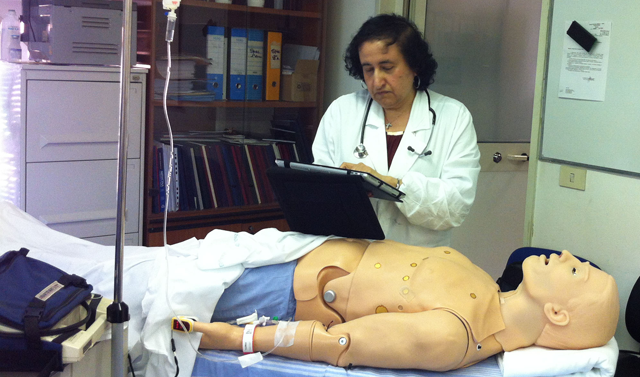}}
\subfigure{\includegraphics[width=0.49\columnwidth]{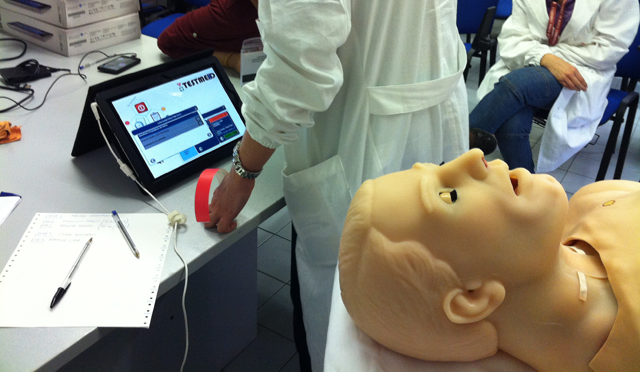}}
\caption{TESTMED system being used by a doctor while attending to a patient (here a patient simulator) in a ward.}
\label{fig:suppainaction}
\end{figure}

The first usability evaluation was conducted in the ward of DEA (Department of Emergency and Admissions) of Policlinico Umberto I in Rome. Figure \ref{fig:suppainaction} shows the TESTMED system being used by a doctor on a patient simulator. There were 7 participants consisting of: 2 doctors and 5 healthcare practitioners. The participants were presented with a patient simulator assumed to be suffering from chest pain problems (see Figure \ref{fig:suppainaction}). They were requested to attend to the patient (patient simulator) with the support of our system.

The participants were also given a questionnaire in order to collect information regarding their background and their assessment of the usability of the system (such as interaction modalities, error management, learnability, effectiveness, etc). Specifically, the questionnaire was composed by 11 statements and a 5 points Likert scale (that ranged from \emph{1 - strongly disagree} to \emph{5 - strongly agree}) that allowed users to express their agreement/disagreement with the statements:
%
\begin{itemize}
\item[Q1] I have a good experience in the use of mobile devices.
\item[Q2] The interaction with the system does not require any special learning ability.
\item[Q3] I judge the interaction with the touch interface very satisfying.
\item[Q4] I judge the interaction with the vocal interface very satisfying.
\item[Q5] I think that the ability of interacting with the system through the touch interface or through the vocal interface is very useful.
\item[Q6] The system can be used by non-expert users in the use of mobile devices.
\item[Q7] The system allows to constantly monitor the status of clinical activities.
\item[Q8] The system correctly drives the clinicians in the performance of clinical activities.
\item[Q9] The doctor may - at any time - access to data and information relevant to a specific clinical activity.
\item[Q10] The system is robust with respect to errors.
\item[Q11] I think that the use of the system could facilitate the work of a doctor in the execution of its activities.
\end{itemize}

The results of the first evaluation are collected in Table~\ref{1_eval_results_table}. On average, the participants indicated a high level of agreement that the prototype could facilitate doctors' work, correctly guided doctors in the execution of CGs, provided doctors with access to information at any time, and enabled doctors to constantly monitor the status of clinical activities.

\renewcommand{\tabcolsep}{1mm}
\begin{table}[h]\scriptsize
\centering
\caption {Results of the first usability test.}
\label{1_eval_results_table}
\begin{tabular}{lccccccccccc}
\hline
\multicolumn{1}{c}{}&\textbf{Q1}&\textbf{Q2}&\textbf{Q3}&\textbf{Q4}&\textbf{Q5}&\textbf{Q6}&\textbf{Q7}&\textbf{Q8}&
\textbf{Q9}&\textbf{Q10}&\textbf{Q11}\\
\hline
\hline
\textbf{User1} & 4 & 3 & 4 & 3 & 2 & 3 & 4 & 4 & 4 & 3 & 3 \\
\textbf{User2} & 4 & 3 & 4 & 2 & 4 & 2 & 2 & 3 & 2 & 3 & 3 \\
\textbf{User3} & 5 & 3 & 4 & 3 & 5 & 2 & 5 & 4 & 5 & 4 & 4 \\
\textbf{User4} & 4 & 4 & 4 & 3 & 3 & 4 & 4 & 4 & 3 & 3 & 4 \\
\textbf{User5} & 3 & 3 & 4 & 2 & 3 & 4 & 4 & 4 & 4 & 3 & 4 \\
\textbf{User6} & 3 & 4 & 5 & 3 & 3 & 5 & 5 & 4 & 4 & 4 & 4 \\
\textbf{User7} & 3 & 4 & 4 & 3 & 4 & 4 & 5 & 4 & 4 & 4 & 4 \\
\hline
\textbf{Avg} & 3,71 & 3,43 & 4,14 & 2,7 & 3,43 & 3,43 & 4,14 & 3,86 & 3,71 & 3,43 & 3,71 \\
\hline
\end{tabular}
\end{table}
On average, the participants fairly agreed that the prototype supported learnability, error management, users who are not experts in using mobile devices. Moreover, the participants emphatically acknowledged the importance of supporting interaction that is less physically and visually demanding. This emphasized the appropriateness of vocal interaction. The participants even requested to provide more support for vocal interaction. Notwithstanding the foregoing, the participants still appreciated the possibility to interact via multi-touch. They also emphasized the need for flexibility by giving users the option to choose, if they so wished, which modality to interact through.

\begin{figure}[h!]
\centering
\includegraphics[width=0.9\columnwidth]{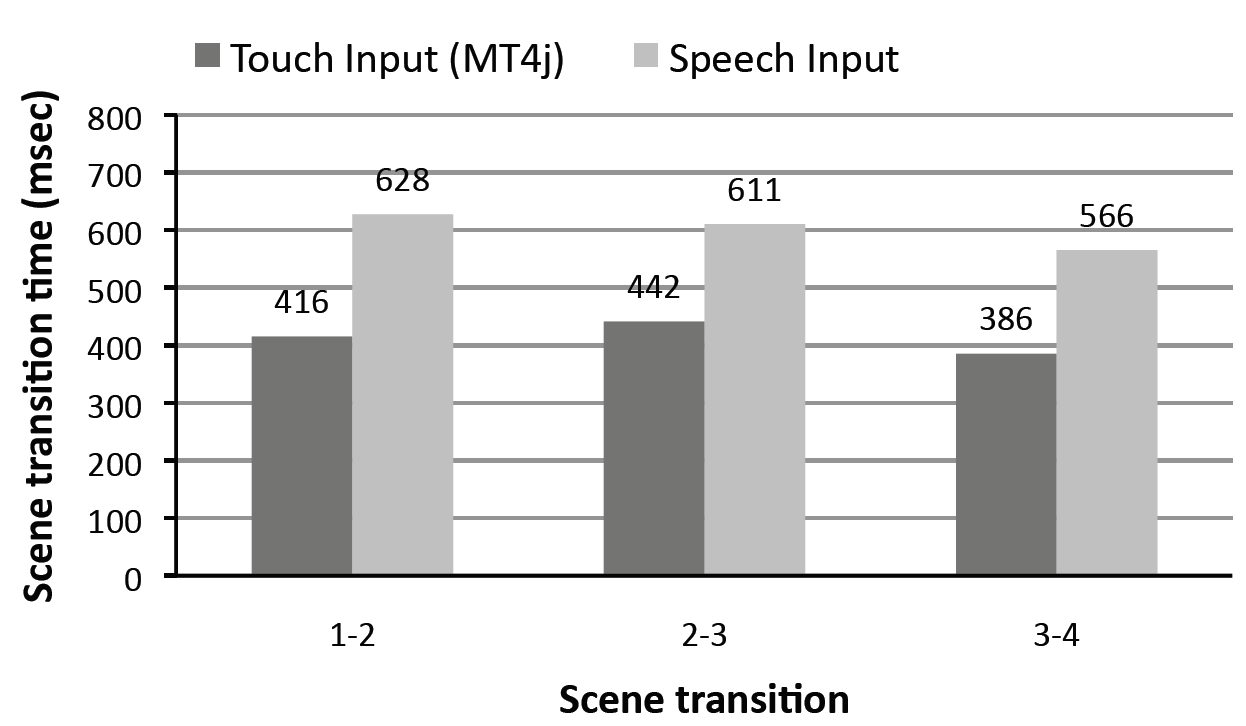}
\caption{The vocal/touch user interface responsiveness tests}
\label{fig:GUITEST}
\vspace{-0.5cm}
\end{figure}

In order to better investigate the responsiveness of the user interface of the first prototype, we carried out further tests for measuring the required time needed for passing from a scene to the following one. A transition between two scenes takes place when a doctor answers to one of the questions of the survey related to the CG deployed into the system.  We repeated the same test twice by using first the touch interface and then the vocal interface. Figure~\ref{fig:GUITEST} shows, on the x-axis, which transition is involved in the current measurement, and on the y-axis the time needed for the generation and the visualization on the screen of the new scene. Since the chest pain score involves 4 different characteristics to be analyzed, 3 scene transitions are required before the generation of the care process. We performed such tests with an ACER Iconia Tab W500 running Windows 7 and provided with 1Ghz AMD CPU and 2 GHz of RAM. On average, about 400 ms were required for the scene transitions when using the touch interface and 6-700 ms for the vocal interface. The key aspect that determines such a delay when using the vocal interface lies in the extra time needed (about 200-250 ms) for contacting the ASR engine. While a delay in speech processing was expected, it is worth noting that it does not significantly impact on system responsiveness and usability, as the overall transition time is lower than 700 ms. The initial prototype was consequently refined based on the results of the first evaluation in order to realize the second prototype.

The second usability evaluation was conducted on the second prototype. In this case, we performed a usability test of the system in the ward of DEA of Policlinico Umberto I in Rome with another set of 7 participants (different users than in the first test) consisting of: 1 doctor, 2 healthcare practitioners, and 4 postgraduate students in medicine. During this usability test, we deployed the chest pain CG into the system. The participants were requested again to attend the patient simulator (see Figure \ref{fig:suppainaction}) with the support of our system. The participants were also required to complete the same survey discussed above for assessing the usability of the system. The results of the second usability evaluation are collected in Table~\ref{2_eval_results_table}.

\renewcommand{\tabcolsep}{1mm}
\begin{table}[h!]\scriptsize
\centering
\caption {Results of the second usability test.}
\label{2_eval_results_table}
\begin{tabular}{lccccccccccc}
\hline
\multicolumn{1}{c}{}&\textbf{Q1}&\textbf{Q2}&\textbf{Q3}&\textbf{Q4}&\textbf{Q5}&\textbf{Q6}&\textbf{Q7}&\textbf{Q8}&
\textbf{Q9}&\textbf{Q10}&\textbf{Q11}\\
\hline
\hline
\textbf{User1} & 4	& 4	& 4	& 4	& 5	& 3	& 4	& 4	& 4	& 4	& 4 \\
\textbf{User2} & 4	& 4 & 5 & 3	& 5	& 2	& 3	& 4	& 2	& 4	& 3 \\
\textbf{User3} & 3	& 3	& 3	& 4	& 4	& 3	& 4	& 4	& 3	& 3	& 4 \\
\textbf{User4} & 5	& 4	& 3	& 2	& 4	& 5	& 5	& 4	& 4	& 5	& 4 \\
\textbf{User5} & 1	& 5	& 5	& 5	& 5	& 5	& 5	& 5	& 5	& 5	& 5 \\
\textbf{User6} & 3	& 5	& 4	& 5	& 4	& 5	& 4	& 4	& 4	& 5	& 5 \\
\textbf{User7} & 3	& 4	& 5	& 5	& 4	& 4	& 5	& 4	& 5	& 4	& 4 \\
\hline
\textbf{Avg} & 3,29	& 4,16 & 4,14 & 4 & 4,43 & 3,86 & 4,29 & 4,14 & 3,86 & 4,29 & 4,14 \\
\hline
\end{tabular}
\end{table}
On average, the participants indicated a high level of agreement that the prototype could facilitate doctors' work, correctly guided doctors in the execution of clinical procedures, provided doctors with access to information at any time, supported learnability, enabled doctors to constantly monitor the status of clinical activities, supported error management, and could be used by users who are not experts in using mobile devices. Moreover, the participants on average acknowledged that the prototype's support for vocal interaction and also for multi-touch was good.

\begin{figure}[h!]
\centering
\includegraphics[width=0.99\columnwidth]{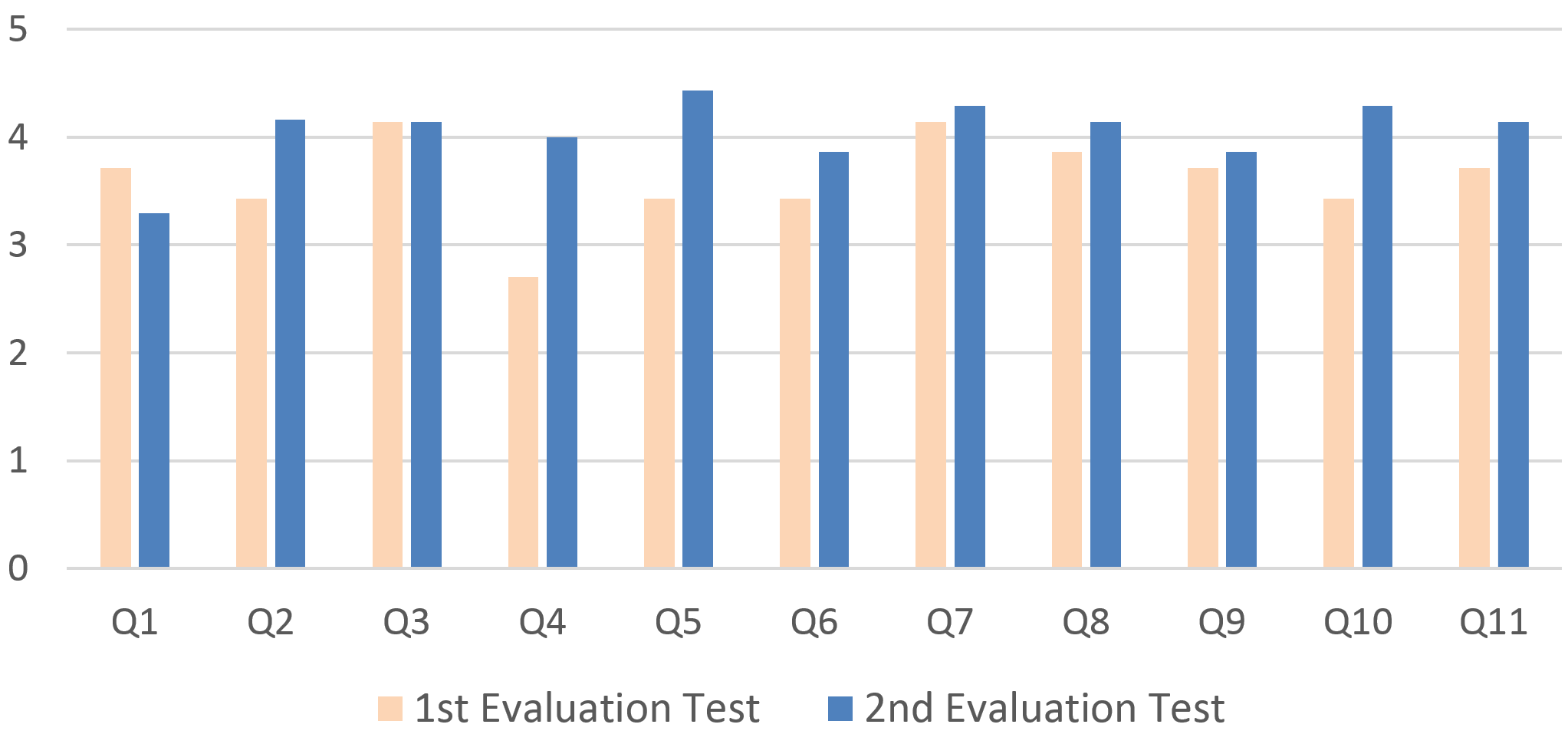}
\caption{Comparison between the questionnaire results of the two usability tests performed.}
\label{fig:comparison}
\end{figure}

It is worth noting that the usability ratings of the second prototype increased tremendously, as shown in Figure~\ref{fig:comparison}, where we compare the results of the two evaluation tests on the basis of the average score for each statement. The second design prototype was considered to have had made great strides toward complying with typical design guidelines for multimodal user interface design~\cite{Laxmisan2007}, for instance our design prototype supports: error management, system feedback (e.g., by enabling doctors to constantly monitor the status of clinical activities), multimodal input and output (e.g., by reducing cognitive and physical demands on the user by supporting eyes-free and hands-free interaction, offering users the option to choose modalities), etc.

\section{Related Work}
\label{sec:related_work}

\subsection{Process-oriented Healthcare Systems}
\label{subsec:healthcare_systems}

In spite of several success stories on the uptake of PAISs in industry and the growing process-orientation of several real world settings \cite{de2010process,ACMTIST2017,AICOMM2018}, BPM technologies have not had the widespread adoption in the healthcare domain yet \cite{Lenz2007,IJBPIM2013}. A major reason for this has been the rigidity enforced by PAISs, which inhibits the ability of a hospital to respond to process changes and exceptional situations in a flexible way \cite{dadam2000clinical}. According to \cite{reichert2011}, whereas efforts were taken to improve and
automate the flow of healthcare processes, their support craves for advanced BPM techniques enabling flexible and adaptive management of such processes.

In this direction, the work~\cite{mansflexibility} identifies the different flexibility requirements related to the application of process technology in healthcare. Although they consider the diagnostic steps of a gynecological oncology process and implement them in four process management systems, the discussion and evaluation focuses on organizational processes and their flexibility requirements.
In~\cite{bpmnpathology}, the authors use the BPMN language for modeling pathology processes for programmed surgical interventions. The proposed models focus on the activities to be performed by different practitioners (including surgeons, nurses, attendants and other pathology department personnel) before, during and after a surgical intervention, capturing the organizational knowledge for coordinating the involved healthcare professionals. However, the resulting models are mainly used for process understanding and improvement, and process enactment through a supporting system is not considered. BPMN is also used in~\cite{bpmnihe} for the definition and analysis of clinical processes related to the tracking of a patient through a healthcare facility from admission to discharge.

A broader perspective on the use of BPMN in healthcare is provided in~\cite{bpmnhealthcare}, where the challenges related to the definition of healthcare processes are considered, including the multi-disciplinary nature of the processes, the flexibility and variability of the involved activities and the interoperability requirements for multiple information systems. Interoperability, application integration and service coordination as starting points for organizational healthcare support are also discussed in~\cite{Lenz2007,processinteroperability}. The authors observe that existing information systems deployed in healthcare settings include many different departmental systems that operate independently, and therefore the computerized support of cross-departmental organizational processes can be related to the problem of data and application integration. While different HL7 standards have been introduced to mitigate the system integration problem, functional integration (i.e., the cooperation of functions of different software components) has not been fully addressed.

In \cite{mans2015process}, the authors investigate how the use of process mining techniques can allow exploiting the event data present in today's EHR systems and address several challenges related to efficiency and costs of managing healthcare processes. According to \cite{manifesto} process mining can be divided into three main branches: \myi process discovery \cite{DBLP:journals/tkde/AalstWM04,mining2011discovery,TKDE}, \myii conformance checking \cite{conf_check_1,conf_check_2,conf_check_3,conf_check_4} and \myiii process enhancement \cite{predicition,fahland2012repairing}.
Specifically, the authors argue that to be able to improve healthcare processes it is important to understand what is really happening during the enactment of care procedures (process discovery) and analyze deviations from the expected or normative process model (conformance checking). Moreover, using the timestamps of events one can identify and diagnose bottlenecks and other inefficiencies (enhancement).


Assuming a service-oriented environment and exploiting the Web services technology, some proposed solutions tackle the issue of enacting healthcare processes through the definition of service orchestration specifications as BPEL processes. In~\cite{bpelmodeldriven}, for example, the authors propose a semi-automatic model-driven approach for the creation of Web service orchestration specifications in BPEL, focusing on an administrative workflow that covers patient admission/discharge/transfer and the scheduling of medical examinations. Process technology, Web services and service-oriented integration are also proposed in~\cite{webservicesemergency} for the automation of inter-organizational emergency healthcare processes, and the approach has recently evolved towards a cloud-based architecture and the use of mobile computing~\cite{cloudemergency}.
In~\cite{yawlserviceflow} the design and implementation of Serviceflow Management System is presented. The system supports the overall care delivery process for the management of acute and chronic care that involves different organizational units. According to a three level architecture, each unit internally manages its own processes and publishes parts of them as services to allow communication with other units; on top of this service level, the overall healthcare process is modeled as a serviceflow that coordinates the available services. 
Another approach that aims at supporting healthcare processes through a service-oriented architecture is presented in~\cite{spdflow}. The authors focus on the procedures of sterile processing departments and identify the main architectural requirements for a workflow system able to manage and automate the work practices. To enable heterogeneous information sharing, integrate different systems and services, and handle failures and exceptions, a service-oriented architecture for the system is proposed; the architecture has been implemented in a prototype system and validated in a decontamination working area.


\subsection{Mobile and multimodal interaction in the healthcare domain}
\label{subsec:mobile_interaction}


Mobile and multimodal user interaction has a long story of success in many real-world settings, including emergency management \cite{capata2008geo,Humayoun09,marrella2011collaboration}, smart environments \cite{lopez2010multimodal,bongartz2012adaptive}, cultural heritage \cite{yang1999smart,collerton2017route}, and - of course - healthcare \cite{chatterjee2009examining}.
When it is administered correctly, this technology can elevate patient care, expand the positive use of mobile devices in hospitals and even ensure healthcare has a more personalised approach.

In this direction, Flood et al.'s work in~\cite{Flood2012} proposes a method for use by application designers during mobile application development in the medical domain to estimate when cognitive overload will occur and can redesign the interface if necessary. The method proposed by Flood et al. therefore targets mobile application designers and developers in the design of user interfaces for the medical domain. While acknowledging that the effort intends to estimate cognitive overload, the effort does not focus directly on design techniques for addressing the problem (such as through multimodal interaction). Jourde et al.'s work in~\cite{Jourde2008} seeks to develop a user interface specification for a multimodal collaborative system for use in a hospital setting. The effort by Jourde et al. is therefore appropriate for application designers. Moreover, both works by Jourde et al.~\cite{Jourde2008} and Flood et al.~\cite{Flood2012} do not directly look into execution of clinical guidelines. Another effort related to our work is HECTOR~\cite{McGee2007}, which is a handheld computer system that was developed to support organizational audit and clinical handover within hospital emergency care teams. HECTOR therefore supports mobile interaction and supports medical staff in handover procedures. It is also worth highlighting the GuideView system which was originally developed as a system aimed at enabling astronauts, who are not necessarily medical experts, to provide medical support for themselves and each other during space exploration missions, when assistance from earth-based medical experts is impractical~\cite{Iyengar2008,Iyengar2009}. It is worth noting that the GuideView system intends to support non-medical experts.

As we noted earlier in Section \ref{sec:introduction}, our work seeks to meet the following requirements: ensure continuity of information flow by supporting mobile access to information, provide support to doctors in the execution of clinical guidelines, and support mobile multimodal interaction in order to reduce the cognitive and physical burden on the doctors. Existing efforts such as the aforementioned ones have either primarily focused on exclusively one requirement, or a partial combination of them. 

\section{Conclusion}
\label{sec:conclusion}

In this paper we have presented the main outcomes of the TESTMED project, aiming at studying and developing
a clinical PAIS supporting doctors during the enactment of CGs in hospital wards, through the interplay of advanced user interfaces deployed on mobile devices and based an integrated speech-recognition, speech-synthesis, multi-touch, and visual interaction framework. The developed system has been evaluated in a real clinical setting through the diagnostic and treatment process foreseen by the chest pain guideline. The results of the user evaluation suggest a good usability of the system and appreciation among medical staff.

Future works include the support for additional guidelines and a new user evaluation targeted at investigating the system's effectiveness towards alleviating cognitive and physical demands on doctors. A further major future challenge consists of determining quantitative and qualitative indicators that, in the long run, can enable us to understand and measure the impact of the system on the overall clinical decision making and collaboration process. 

\bibliographystyle{spbasic}      

\bibliography{biblio}   

\end{sloppypar}
\end{document}